\documentclass[aps,prd,twocolumn,superscriptaddress,nofootinbib,amssymb]{revtex4}
\pdfoutput=1

\usepackage{amsmath,amssymb,amsfonts,bbold}
\usepackage{psfrag}
\usepackage{url}
\usepackage{color}
\definecolor{darkblue}{rgb}{0.1,0.1,.7}
\definecolor{purple}{rgb}{0.6,0,0.6}
\definecolor{orange}{rgb}{0.9,0.6,0}
\usepackage{pgfplots}
\pgfplotsset{compat=1.6}
\usepackage{tikz}
\usetikzlibrary{shadings}
\usepgfplotslibrary{patchplots}
 
\newcommand{\bea}{\begin{eqnarray}}
\newcommand{\eea}{\end{eqnarray}}

\def\li{|} 
\def\ri{|} 
\def\la{\langle}
\def\ra{\rangle}

\newcommand {\bes} {\begin {equation*}}
\newcommand {\ees} {\end {equation*}}

\def\is{\! & \! = \! & \!}

\def\spc{\hspace{1pt}}
\usetikzlibrary{calc,3d}
\usetikzlibrary{decorations.markings}
\usetikzlibrary{decorations.pathmorphing}
\def\tr{{\rm tr}}
\usepackage{verbatim}
 \def\smpc{\hspace{.5pt}}
\def\nspc{\hspace{-1pt}}

\def\be{\begin{equation}}
\def\ee{\end{equation}}

\begin{document}

\title{ER\,\spc =\,\spc EPR revisited: \\[1.6mm]
On the Entropy of an Einstein-Rosen Bridge}

\def\spc{\hspace{.5pt}}

\author{Herman Verlinde}
\affiliation{Department of Physics, Princeton University, Princeton, NJ 08544, USA}

\date{\today}

\begin{abstract}
We propose a new link between entropy and area:  an eternal black hole with an ER bridge with cross-section $A$ can carry a macroscopic amount of quantum information, or be in a mixed state, with entropy bounded by $S \leq A/4G_N$.  We substantiate our proposal in the context of AdS$_{3}$ and JT gravity, by using the Island prescription and replica wormhole method for computing the black hole entropy.  We argue that the typical mixed state of a two sided black hole takes the form of an entangled  `thermo-mixed double' state with only classical correlations between the two sides. 
Our result for the von Neumann entropy of a post-Page time two-sided black hole is smaller by a factor of two from previous answers.  
Our reasoning implies that black hole quantum information is topologically protected, similar to the information stored inside a topological quantum memory. 
\vspace{-2.5mm}

\end{abstract}

\def\calO{{b}}
\def\be{\begin{equation}}
\def\ee{\end{equation}}

\setcounter{tocdepth}{2}
\maketitle
\subsection*{Introduction}
\vspace{-3mm}

The Bekenstein-Hawking relation \cite{bh} between the entropy and horizon area $A$ of a black hole
\bea
\label{bh}
S \is \mbox{\it A}/{4G_N}
\eea
forms one of the central guiding principles in the search for a quantum theory of gravity. The precise meaning and realm of applicability of the relation \eqref{bh}, however, are still only partly understood. The two most well-formulated interpretations of equation \eqref{bh} are that it quantifies:\\[1.3mm]
i) {\it the total number of quantum mechanical micro-states of a black hole with specified macroscopic properties,} or\\[1.3mm]
ii) {\it the amount of entanglement across the event horizon connecting the two sides of an eternal black hole.}\\[1.3mm]
Both interpretations are well founded. The first applies to one sided black holes and has found confirmation in supersymmetric string theory and in AdS/CFT. It states that black holes are not all identical but store quantum information, encoded in subtle properties of the microscopic quantum state.  The second interpretation finds support within AdS/CFT via the Ryu-Takayanagi formula \cite{RT} and the proposed holographic interpretation of the thermofield double 
\bea
\label{tfd}
\li \mbox{\small \rm TFD}\ra \is \sum_n \, \sqrt{p_n\!} \; \li n \ra_{\! L}\spc \li n \ra_{\! R} 
\eea
with $p_n = {{e^{-\beta E_n}}}/{Z}$ as the state of a two sided black hole  with an Einstein-Rosen bridge \cite{Maldacena:2001kr}.  This presumed linkage between entanglement and space-time connectedness \cite{vanraamsdonk}\cite{EH}\cite{ER=EPR} now goes by the slogan ER\,=\,EPR.

The two statements i) and ii) seem to refer to a different notion of entropy. However, they are directly related. Given i), it is natural to consider the one sided black hole in a thermal mixed state 
\bea
\label{mixed}
\rho_R
 \is  
 \sum_n \, p_n \, 
 \li n\ra_R \la n \ri  \\[-5mm]\nonumber
\eea
with von Neumann entropy\\[-4mm]
\bea
\label{sright}
S_{R}\spc \is \spc -{\rm tr}(\rho_{\mbox{\tiny $R$}}\! \log \rho_{\mbox{\tiny $R$}}) \spc = \spc -\sum \spc p_n \log p_n\spc = \spc S_{BH}.
\eea
We can formally identify $\rho_R$ with the reduced density matrix of the right-hand side of the thermofield double \eqref{tfd} obtained by tracing out the left-hand side
\bea
\label{rhor}
\rho_R \is \tr_{L}  \bigl(\spc \rho_{\spc {\rm TFD}} \spc \bigr)\, \qquad 
\rho_{\smpc {\rm TFD}} =\li\mbox{\small \rm TFD}\ra\la  \mbox{\small \rm TFD}\ri
\eea
The TFD represents the formal purification of the thermal mixed state of a one-sided black hole.

Statement ii) posits that this formal purification describes the  extended space-time of a two-sided  black hole with an ER bridge.
This identification finds support within the idealized context of AdS/CFT. From a physical perspective, however, there is something unsatisfactory to the claim that a two-sided black hole is described by the TFD. The TFD is an artificial theoretical construct: it's a {\it unique} state with {\it zero} von Neumann entropy.
A black hole, on the other hand, is a macroscopic object, which under normal circumstances would have a large entropy, quantifying the number of microstates with the same macroscopic characteristics, the maximal amount of entanglement the object can have with its environment,  or the maximal amount of quantum information it can contain. Indeed, as soon as a  black hole is embedded inside an ambient space time,  there is no reason to believe that it will stay in a pure state.

This begs the question: what is the maximal entropy, or amount of quantum information, that a two sided black hole can contain? How much entanglement can it have with its environment? Below we will argue that\\[1.5mm]
iii) {\it a space-time with an ER bridge with cross sectional area $A$ can carry a macroscopic amount of quantum information with entropy equal to $S_{BH} = A/4G_N$.}\\[1.5mm]
Since any macroscopic object usually decoheres into a mixed state, a corollary of this assertion is that\\[1.5mm]
iii') {\it the local space-time region of an ER bridge is typically in a mixed state with entropy $S_{BH} = A/4G_N$.}\\[1.5mm]
We propose that these two statements, together with~the accepted insight ii),
form the three basic holographic entropy relations associated with an ER bridge. 

We give two arguments in support of this new holographic principle. The first is based on the Island prescription for determining the entanglement wedge of the emitted radiation of a black hole \cite{Penington:2019npb}-\cite{replica2}. For a two-sided black hole, the Island prescription was examined in \cite{replica2}. It was argued that the dominant gravitational saddle points that contribute to the n-th Renyi entropy include replica wormholes connecting the $n$ replica space-times. We will show, via a direct calculation in JT gravity and pure 3D AdS gravity, that this replica wormhole prescription implies that the two-sided black hole state takes the form of a special entangled mixed state of the form (here $p_n = e^{-\beta E_n}/{{Z}}$) 
\bea
\label{tmdt}
\rho_{\smpc {\rm TMD}} \, = \,  \sum_n \,p_n\,  
 \li n \ra_L 
\la n\ri \otimes \spc \li n \ra_R \la n\ri  
\eea
with entropy $S_{BH} = A/4G_N$. We will call \eqref{tmdt} the `thermo-mixed double', as it describes the dephased version of the thermofield double.

The second argument in support of our proposal relies on AdS/CFT and the insight that space-time behaves like a quantum error correcting code \cite{VV-QEC}-\cite{VJ}. i.e. the effective bulk QFT lives within a small code subspace  of the full microscopic Hilbert space \cite{EWR}. The holographic rule iii) then arises from the observation that the Poincar\'e recurrence time as seen from within the code subspace is much shorter than the Poincar\'e recurrence time of the microscopic theory. The huge hierarchy between the two time scales means that the true microscopic age of the two sided black hole is kept hidden from the low energy observer. This lack of knowledge of the black holes age is quantified by the Bekenstein-Hawking entropy $S_{BH}$.

 \vspace{-3mm}

\subsection*{Three thermal states}

\vspace{-3mm}

Here we propose a geometric definition of the thermomixed double state \eqref{tmdt} and compare some of its properties with those of two other thermal states.

Recall the geometric argument for why the TFD state describes a two-sided black hole \cite{Maldacena:2001kr}. By formally identifying the left Hilbert space ${\cal H}_L$ spanned by $\li n\ra_L$ with the dual ${\cal H}^*_R$ of the right Hilbert space spanned by $\li n \ra_R$, we can rewrite the TFD as the square root of a thermal density matrix. The TFD can thus be thought of as prepared via a euclidean path integral over half a thermal circle. Via the AdS/CFT dictionary, the associated dominant saddle point in the gravity dual is a euclidean geometry with this half circle as its boundary. We thus may graphically represent the TFD  via\\[-4mm]
\bea
\li \mbox{\small \rm TFD}\ra\; \is\; \raisebox{-4mm}{\includegraphics[scale=0.45]{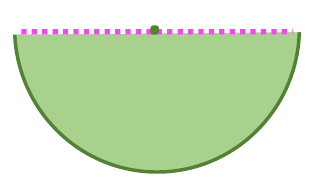}}\ \
\eea
The boundary represents the half thermal circle, the solid region describes the lower half of a euclidean two-sided black hole geometry, and the dashed line is the bulk state.  After Wick rotating, the point at the center becomes the event horizon of the lorentzian two sided black hole \cite{Maldacena:2001kr}.

Using this geometric notation, we represent the thermal density matrix $\rho_R$, obtained by tracing out the left region, by\\[-6mm]
\bea
\label{rhord}
\rho_R \ \is \  \raisebox{-7mm}{\includegraphics[scale=.44]{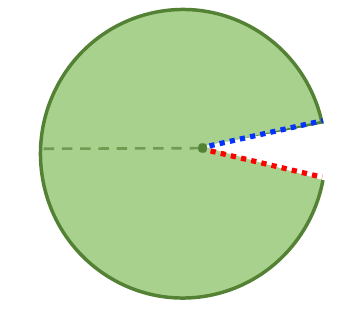}}\\[-3mm]\nonumber
\eea
This figure can be thought of as obtained by first drawing the total density matrix $\rho_{\smpc {\rm TFD}}$ of the TFD state as two half discs.
Tracing out the left CFT corresponds to gluing together the left-halves of the two half discs, thereby forming the `pacman'  figure \eqref{rhord}. The center point at the corner of the opening indicates the horizon and the end-point of the entanglement wedge of the right CFT.

Our proposed geometric representation of the gravitational saddle point associated with the thermo-mixed double state \eqref{tmdt} looks like two half disks connected via an `Island' region in the middle \\[-4.5mm]
\bea
\label{tmdsaddle}
\rho_{\smpc {\rm TMD}} \ \is \raisebox{-6.5mm}{\includegraphics[scale=.465]{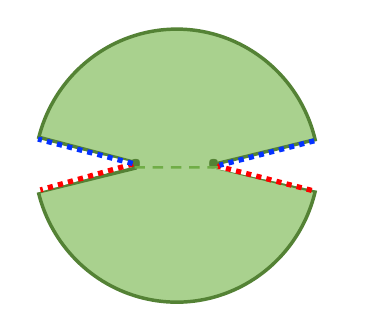}}\\[-3.5mm]\nonumber
\eea
This `janus pacman' figure represents a mixed density matrix on the tensor product of two holographic CFTs.  The Island region is the geometric remnant of the entanglement with the environment ${\cal E}$. Via the Island prescription \cite{Penington:2019npb}\cite{Almheiri:2019psf}, the entanglement wedge of the external radiation contains an interior region and thus tracing out ${\cal E}$ in effect glues together the two geometries along this Island region. We will substantiate the above geometric representation of the TMD later on via a gravity computation of the Reny\'i entropies.

For comparison, let us also introduce the factorized thermal mixed state
\bea
\label{rhof}
\rho_L \otimes \rho_R \; \is \; 
 \sum_n \, p_n \, 
 \li n\ra_L \la n \ri\, 
 \otimes 
 \, \sum_{\tilde n} \, p_{\tilde n} \, 
 \li \tilde{n}\ra_R \la \tilde{n} \ri \ \ 
\eea
representing the maximally mixed state of the two CFTs.
The corresponding saddle point geometry factorizes into two disconnected pacman discs\\[-4.5mm]
\bea
\rho_L \otimes \rho_R \ \is \ \; \raisebox{-6.5mm}{\includegraphics[scale=.43]{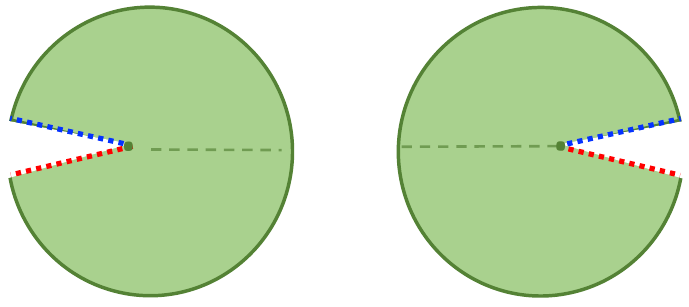}}\quad \\[-3mm]\nonumber
\eea

The TMD, TFD  and factorized mixed state $\rho_L \otimes \rho_R$ all give rise to the same reduced density matrices on each side. E.g. the right density matrix 
\bea
\label{rhor}
\rho_R \is  \tr_{L}\nspc  \bigl( \rho_{\spc {\rm TMD}} \bigr) =  \tr_{L}\nspc  \bigl(\rho_{\spc {\rm TFD}}  \bigr) = \tr_{L}\nspc(\rho_L \otimes \rho_R)\
\eea
equals the thermal state \eqref{rhor} with non Neumann entropy \eqref{sright}
equal to  $S_{BH}$. So when viewed from one side, the three  states are indistinguishable. 

The difference shows up in the von Neumann entropy $S(\rho) = - \tr(\rho\log\rho)$ of the combined $L\cup R$ density matrix. Evidently,  $S(\rho_{\rm TFD})=0$ for the thermofield double state while  $S(\rho_L\otimes \rho_R) = 2S_{BH}$ for the factorized thermal  state. The entropy of the TMD state, on the other hand, equals $S(\rho_{\rm TMD}) =S_{BH}$.  The mutual information $I_{LR} \, = \, {S_L + S_R - S_{LR} }$ between the left and right regions for each case thus equals
\bea
I_{LR}\,\is \, \left\{ \begin{array} {c} {2S_{BH}\  \qquad \qquad  \mbox{for\, \small TFD}\ \ }\\[2mm] {\ \, {S_{BH}}\  \qquad \qquad  \, \mbox{for  \,\small TMD}\ \ }
\\[2mm]
{\ \ \; 0\  \qquad \qquad \quad \    \mbox{for  $\rho_L\! \otimes\!\spc \rho_R$}}\end{array}\right.\qquad
\eea
We observe a clear distinction. In the factorized thermal state, the two sides do not share any mutual information.
The relative factor of two in the mutual information between the TFD and TMD state makes an essential difference:  in the TFD state, the two sides share  true quantum mutual information in the form of quantum entanglement; in the TMD state, on the other hand, the two sides share only classical Shannon mutual information. In other words, the TMD state exhibits only classical correlations and not true quantum entanglement.

\vspace{-3mm}

\subsection*{Global geometry of an ER bridge}

\vspace{-2mm}

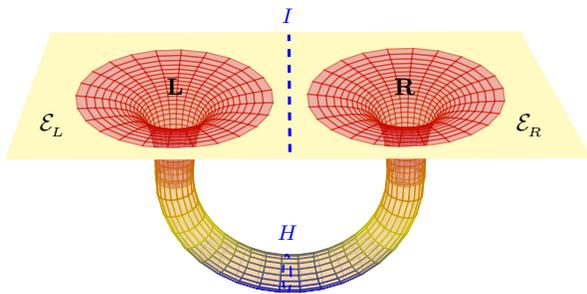
\begin{figure}[t]
\begin{center}
\vspace{-1.3cm}
\begin{tikzpicture}[scale=1.3]
\coordinate (P1) at (3cm,9.75cm); 
    \coordinate (A1) at (0.66cm,3.25cm); 
	\coordinate (A2) at (6.66cm,3.265cm); 
	\coordinate (A3) at ($(P1)!.8!(A2)$); 
	\coordinate (A4) at ($(P1)!.8!(A1)$);
    \fill[yellow!30] (A1) -- (A2) -- (A3) -- (A4) -- cycle; 
    \begin{axis}[hide axis,axis equal,rotate around z=-9] 
 \addplot3[opacity=.2, domain=0:360,y domain=-180:0, samples=20,surf,z buffer=sort,opacity=.8,color=purple!40!yellow!40!]
            ({.832*(6.3+(6.19+ 1.0 * cos(x)) * cos(y)) - .55 * sin(x)},{.832 * sin(x) + .55*(6.31 +(6.18+ 1.0 * cos(x)) * cos(y))},
            {-1+.95(6.1+ 1.0 * cos(x)) * sin(y)});
            \addplot3[opacity=.5, color=blue!30!,surf,domain=0:360,y domain=0:2,color=purple!60!]
                ({-.15+(1+.10*y*y)*cosh(y)*cos(x)},{.45+(1+.10*y*y)*cosh(y)*sin(x)},{-.55+ 5.1*tanh(y)});
                  \addplot3[opacity=.5,surf,domain=0:360,y domain=0:2,color=purple!60!]
                ({10+ (1+.1*y*y)* cosh(y)*cos(x)},{7.5+ (1+.1*y*y)*cosh(y)*sin(x)},{-.55+5.1*tanh(y)});   
        \end{axis}
    \fill[yellow!30] (2cm,3.252cm) -- (2.8cm,3.253cm) -- (2.8cm,3.39cm) -- (2.4cm,3.37cm) -- (2cm,3.38cm) -- cycle;
    \fill[yellow!30] (4.3cm,3.261cm) -- (5.1cm,3.262cm) -- (5.1cm,3.40cm) -- (4.7cm,3.38cm) --  (4.3cm,3.40cm) --cycle; 
\draw (2.4,4) node {$\mbox{\bf L}$};
\draw (4.75,4) node {$\mbox{\bf  R}$};
\draw (6.04,3.58) node {$\mbox{$\cal E$}_{\! {}_{{}^R}}$};
\draw (1.16,3.58) node {$\mbox{$\cal  E$}_{\!{}_{{}^L}}$};
\draw [dashed, color = blue,thick] (3.57,3.33) -- (3.56,4.53);
\draw [dashed, color = blue,thick] (3.58,3.33) -- (3.57,4.53);
\draw[color=blue](3.55,2.5) node {$\mbox{\footnotesize $H$}$};
\draw [color=blue,dashed,thin] (3.54,2.08) ellipse (.4mm and 2mm);
\draw [color=blue,dashed,thin] (3.55,2.08) ellipse (.4mm and 2mm);
\draw[color=blue](3.55,4.72) node {$\mbox{\footnotesize $I$}$};
\end{tikzpicture}\qquad
\end{center}
\vspace{-2.5cm}
\caption{Spatial slice of an ER bridge. The two asymptotic regions ${\cal E}_L$ and ${\cal E}_R$ are connected via the interface $I$.}
\vspace{-3mm}
\end{figure}

Before presenting our entropy arguments, it will be useful to briefly describe the classical geometry of the ER-bridge in a bit more detail. There are two basic setups that we will consider. The first situation is
 shown in Fig 1.  It describes two black holes connected via an ER bridge, both embedded within the same space-time. The total space in this case is non-simply connected. 
 
 Let $L$ and $R$ denote the near-horizon regions of each black hole, glued together at the bifurcate horizon $H$ with area $A$. The two regions $L$ and $R$ are surrounded by their respective asymptotically flat regions ${\cal E}_L$ and ${\cal E}_R$. Regions ${\cal E}_L$ and ${\cal E}_R$ are in turn connected via the interface $I$, as indicated. The two sides can also communicate by sending messages via the exterior space-time. The length of the ER bridge is assumed to be much shorter than the outside distance between the two black holes.

The second situation is the usual two-sided AdS black hole shown in figure 2. The ER-bridge in this case interpolates between two otherwise disconnected AdS space-times. In this situation, the only potential means of communication between the two sides is by having two messengers meet inside the black hole interior.

\begin{figure}[t]
\begin{center}
\vspace{-.2cm}
${}$\ \  \includegraphics[scale=0.365]{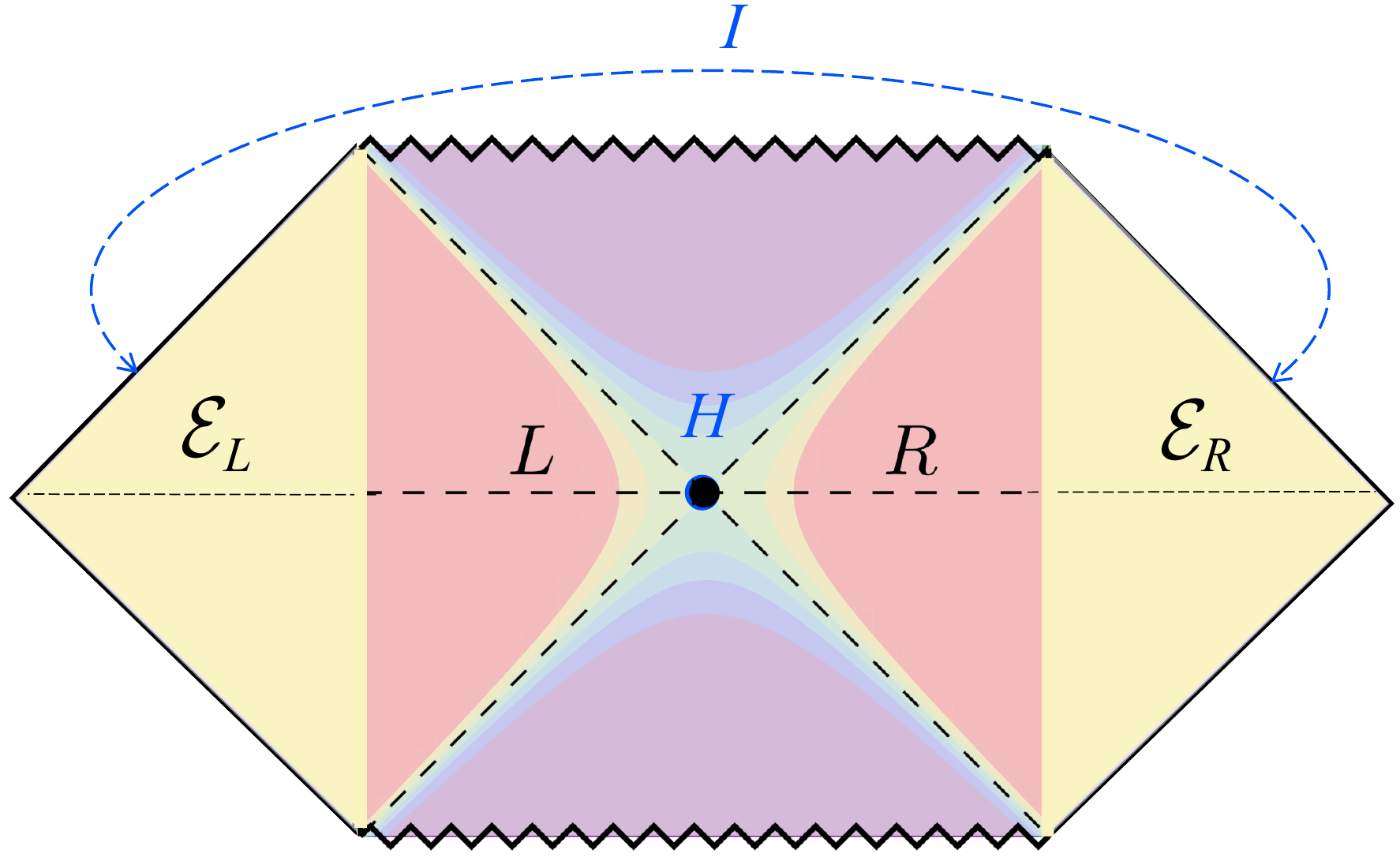}
\end{center}
\vspace{-1mm}
\caption{Penrose diagram of an ER bridge. The time variables defined on the asymptotic regions ${\cal E}_L$ and ${\cal E}_R$ are connected via an identification $I$.}
\vspace{-2mm}
\end{figure}

We assume that the space-time metric of an ER bridge can be chosen to take the quasi-static form
\bea
\label{static}
ds^2_{d+1} \is - f(x)^2 dt^2 + g_{ij}(x) dx^i dx^j.
\eea
Here $f(x)$ vanishes, and thus flips sign, at the~horizon. As a result, the space-time \eqref{static} does not admit a global time-like Killing vector: the locally defined  Killing vector $\frac{\partial\ }{\partial t}$ switches sign when passing through the horizon $H$, and thus comes back in the opposite direction when transported around the non-contractible cycle in Fig 1.

 To write the metric as in \eqref{static}, we must introduce two coordinate charts, or `time zones': a left region that includes $L$ and ${\cal E}_L$ with time coordinate $t_L$, and a right region that combines $R$ and ${\cal E}_R$ with time coordinate $t_R$. The transition functions at the horizon $H$ and interface $I$, that prescribe the local identification between the left and right time coordinates $t_L$ and $t_R$, are given by
\bea
\left. \begin{array}{c} {t_L = t_R  \qquad \qquad  \; \;  \mbox {at \spc $I$}} \\[2mm]
{\, t_L  =  2t-t_R \qquad  \; \mbox{ at $H$}} 
\end{array}\right\} \ \ \Leftrightarrow \ \ t_L = t_R = t
\eea
Here $t$ defines the age of the black hole. These transition functions break global time-translation invariance, and select the time slice $t \!\spc=\!\spc t_L \!\spc =\!\spc t_R\!\spc = 0$ as a special time instant. 
The $t$-dependence is invisible locally, and only noticeable by considering non-local observables that are sensitive to the transition functions at both at $H$ and~$I$. To see this, note that we could make a redefinition $t_L \to 2t-t_L$ of the left time coordinate, after which the transition functions at $H$ and $I$ would take the new form
\bea
\left. \begin{array}{c} {t_L = t_R   \qquad \quad \ \ \;  \; \mbox {at \spc $H$}} \\[2mm]
{\, t_L  =  2t-t_R \;  \qquad  \mbox{at $I$}} \; 
\end{array}\right\} \ \ \Leftrightarrow \ \  t_L = t_R = t
\eea
The time dependence in now encoded in the transition function at the interface $I$ instead of at the horizon $H$. 

The take away lesson of the above discussion is that the two-sided black hole space time is time dependent, but that this $t$ dependence can not be noticed \footnote{In case the two sided black hole undergoes Hawking evaporation,  this would induce a slow adiabatic time-dependence of the external geometry, reflecting the decreasing mass of the black hole. This slow time dependence is encoded in the evolving form of the spatial metric and superimposed on top of the time-dependence encoded in the transition functions between the left and right time coordinates at $H$ and $I$.}  by a single local observer traveling through the horizon, nor by one that stays outside, since neither  can simultaneously probe the two transition functions at $H$ and at $I$. A pair of observers, on the other hand, can together make an attempt to measure the time variable $t$, by each jumping from opposite sides into the black hole, and comparing their watches before and after they jump in.

\vspace{-3mm}

\subsection*{Non-local phases and generalized TFD states}

\vspace{-3mm}

Let us temporarily assume that the near-horizon geometry of the two-sided black hole is perfectly decoupled from its environment and prepared in a pure state, just like the thermofield double state in AdS/CFT. We will still introduce an identification $I$ between the time variables on each side, as indicated in Fig 2. The density matrix on each side takes the form of a thermal mixed state \eqref{rhor}. Given this fact, how can the black hole carry any quantum information? 

A natural way to store quantum information inside the two-sided black hole is by introducing arbitrary phase factors in the eigen basis decomposition of the TFD state 
\bea
\label{tfdalpha}
\li \mbox{\small \sc TFD}\ra\raisebox{-1.5pt}{${\nspc\smpc}_{\bf \alpha}$}\is 
\,\sum_n \; e^{i\alpha_n}\, \sqrt{p_n\!} 
\; \li n \ra_{\! L}\spc \li n \ra_{\! R} 
\eea
These generalized TFD states share most of the properties of the standard TFD state with $\alpha_n=0$. In particular, the phases $e^{i\alpha_n}$ can not be extracted via local measurements in the neighborhood of the horizon, but can only be measured via non-local observables that probe the non-trivial topology of the full spatial slice. 

This interpretation of the $e^{i\alpha_n}$ as topological phase factors is made evident by comparing \eqref{tfdalpha} with the time evolved TFD state
\bea
\li \mbox{\small \sc TFD}(t)\ra
\is 
\sum_n  e^{-2i E_n t }\spc \sqrt{p_n\!} \; 
\li n \ra_{\! L} \li n \ra_{\! R}\, ,
\eea
while noting that the time coordinate $t$ specifies a global boundary condition, that can only be measured via a non-local observable. In fact, assuming that the holographic CFT has a random energy spectrum, it is possible to argue that for any set of phases $e^{i\alpha_n}$, one can always find some special time instant $t_\alpha$ such that 
\bea
\label{talpha}
e^{i(\alpha_n-\alpha_m)} \is e^{-2i (E_n-E_m) t_\alpha} 
\eea 
for all $n$ and $m$. Hence we may identify the generalized TFD state with standard TFD states evolved over the corresponding time $t_\alpha$
\bea
\li \mbox{\small{TFD}}\ra\raisebox{-1.5pt}{${\nspc\smpc}_{\bf \alpha}$} \is \li \mbox{\small{TFD}}(t_\alpha)\ra
\eea
Note that $t_\alpha$ is defined modulo the quantum Poincar\'e recurrence time $t^{\rm cft}_{P}$ of the CFT.  For a generic set of phases $e^{i\alpha_n}$, the special time $t_\alpha$ is double exponentially large in the entropy of the black hole.

Since the  generalized TFD states $\li \mbox{\small \sc TFD}\ra\raisebox{-1.5pt}{${\nspc\smpc}_{\bf \alpha}$}$ all look indistinguishable to any local observer near, inside or outside the horizon, local observers can also not distinguish between the pure TFD state and states described by arbitrary superpositions  
\bea
\label{superp}
|\Psi \rangle \is \sum_{{\bf \alpha}}\, c_\alpha\, \li \mbox{\small{TFD}}\ra\raisebox{-1.5pt}{${\nspc\smpc}_{\bf \alpha}$} \, = \, \sum_{{\bf \alpha}}\, c_\alpha\, \li \mbox{\small{TFD}}(t_\alpha)\ra
\eea
with $\sum_\alpha |c_\alpha|^2=1$. This superposition state looks quite strange from the viewpoint of the {\it global} observer: it describes a coherent superposition of different {\it global} geometries, each given by different time-evolved two-sided black holes. However, without having done the explicit non-local measurement of the global time variable $t=t_\alpha$, the {\it local} observer has no knowledge whatsoever about its value. Moreover, given that the typical times $t_\alpha$ are exceedingly large compared to any reasonable time-scale accessible to a low energy observer, it seems more appropriate to view the phases $e^{i\alpha_n}$ as truly quantum mechanical non-integrable phase factors, analogous to Aharanov-Bohm phases, rather than the specification of a precise classical geometry. It is then reasonable to consider superpositions of states of the form \eqref{superp}.

\vspace{-2mm}

\subsection*{Dephasing and the thermo-mixed double}

\vspace{-3mm}

How much quantum information can a general superposition \eqref{superp} of generalized TFD states carry? To find this out, we first note that the generalized TFD states are all approximately orthogonal 
\bea
\raisebox{-1.5pt}{${}_\beta$}\la \mbox{\small \sc TFD} \li \mbox{\small \sc TFD}\ra\raisebox{-1pt}{${\nspc}_\alpha$} 
\is \sum_n \, p_n\, e^{i (\alpha_n - \beta_n)} \, 
\, \simeq \, \delta_{\alpha\beta}\;  
\eea
up to terms of order $e^{-S_{BH}/2}$, provided we choose the set of phases $e^{i \alpha_n}$ and $e^{i \beta_n}$  sufficiently random. 

Given that the non-local phases are invisible to local observers, it is most fitting to describe the state of the two sided black hole by an {\it incoherent} sum of all basis states  $\li \mbox{\small \sc TFD}\ra\raisebox{-1pt}{${\nspc}_\alpha$}$. A more direct physical reason for adopting this description is that in the situation indicated in Fig 1 and 2, the two sided black hole is embedded inside, and thus highly entangled with, its ambient space-time ${\cal E}$. Since the ambient space-time constitutes a much larger quantum system than the two-sided black hole itself, we should expect that, after we wait long enough,  the black hole will evolve into a maximally mixed state, given by the incoherent sum of all states compatible with the macroscopic properties seen by the low energy observer. 

This leads us to introduce the mixed state given by the incoherent sum over all generalized TFD states
\bea
\label{tmdo}
\rho_{\smpc {\rm TMD}} \, \is \,\frac 1 N\,  \sum_{\mbox{\scriptsize \boldmath$\alpha$}}\, \li \mbox{\small \sc TFD}\ra\raisebox{-1.5pt}{${\nspc}_\alpha$\!} \,\; \raisebox{-1.5pt}{${}_\alpha$\!}\la \mbox{\small \sc TFD} \spc \li 
\eea
Here $N$ counts the total number of basis states.
We will call the state \eqref{tmdo} the thermo-mixed double, since it arises after applying complete dephasing to the thermo-field double state. Using that the sum over the complete set of random phases yields a diagonal distribution
\bea
\frac 1 N\, \sum_{\mbox{\scriptsize \boldmath$\alpha$}}\;  e^{i(\alpha_n - \alpha_m)} \, = \, \delta_{nm}
\eea
we deduce that the thermo-mixed double takes the following more explicit form
\setcounter{equation}{5}
\bea
\label{tmdtt}
\rho_{\smpc {\rm TMD}} \, = \,  \sum_n \,p_n\,  
 \li n \ra_L 
\la n\ri \otimes \spc \li n \ra_R \la n\ri  
\eea
with $p_n = e^{-\beta E_n}/{{Z}}$. This mixed state describes a generic two sided black hole with entropy $S_{BH}$. 

\setcounter{equation}{24}

As explained earlier, the TMD state has only classical correlations across the horizon and does not support any true quantum entanglement. So our proposal,
that \eqref{tmdt} represents the generic state of a two-sided black hole,  seems at odds with the ER = EPR proposition.  Note, however, that after embedding the black hole inside an ambient space time ${\cal E}$, the role of entanglement across the event horizon becomes less clear cut. We can introduce a purification of the TMD of the tripartite form
\bea
\label{psi}
|\Psi\ra \is \sum_n \, \sqrt{p_n\!} \; \, |n\ra_L\, |n\ra_R \, |\psi_n\ra_{\cal E}
\eea
with $ |\psi_n\ra_{\cal E}$ an orthonormal set of states of the environment.
The entanglement between the black hole and the environment is built up through local interactions and the Hawking process.
We can formally split ${\cal E}$ into a left and a right environment ${\cal E}_L$ and ${\cal E}_R$. Assuming that both sides are truly decoupled,  the environment state $\li\psi_n\ra_{\cal E}$ would factorize into a product state $|\phi_n\ra_{{\cal E}_L} | \tilde{\phi}_n\ra_{{\cal E}_R}$. Inserting this back into \eqref{psi}
\bea
\label{psid}
|\Psi\ra \is \sum_n \, \sqrt{p_n\!} \; \, |n\ra_L\, |\phi_n\ra_{{\cal E}_L} \otimes |n\ra_R \,| \tilde{\phi}_n\ra_{{\cal E}_R}
\eea
 would reinstate the quantum entanglement across the horizon. On the other hand, if ${\cal E}_L$ and ${\cal E}_R$ are not decoupled, the quantum entanglement between the two sides dissipates. For the tripartite state \eqref{psi}, this dissipation of entanglement does not completely decohere the TFD state (which would produce the factorized thermal state \eqref{rhof} with entropy $2S_{BH}$) but leads only to dephasing, thereby producing TMD state \eqref{tmdtt} with entropy $S_{BH}$. The classical correlations across the ER bridge remain in place due to a topological protection mechanism. We will make this analogy with topologically protected quantum information more manifest below.

\vspace{-3mm}
\subsection*{Replica wormholes and Reny\'i entropy}
\vspace{-3mm}

Now let us return to the proposed holographic prescription \eqref{tmdsaddle} for obtaining the TMD state. We would like to verify that it indeed gives rise to a mixed state of the precise TMD form \eqref{tmdt}. 

First let us compare some basic properties of  \eqref{tmdsaddle}  and \eqref{tmdt}. Clearly, both states are neither pure, nor factorized. Moreover, both states have the property that if we trace over the left or right CFT, we obtain a thermal density matrix on the remaining CFT. 

Next let us look at the purity of the mixed state, defined by the trace over the square of the $\rho_{\rm TMD}$ density matrix. We claim that the corresponding bulk geometry looks as follows\\[-4.5mm]
\bea
\tr(\rho_{\rm TMD}^{\, 2}) \; \is\; \raisebox{-7mm}{\includegraphics[scale=.52]{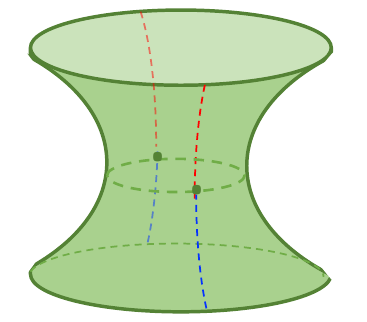}}\\[-2.5mm]\nonumber
\eea
The geometric proof is straightforward:  to obtain the purity, we first have two take two copies of the TMD state  \eqref{tmdsaddle}, and then glue the two together along the corresponding edges as follows\\[-3mm]
\bea
\includegraphics[scale=0.51]{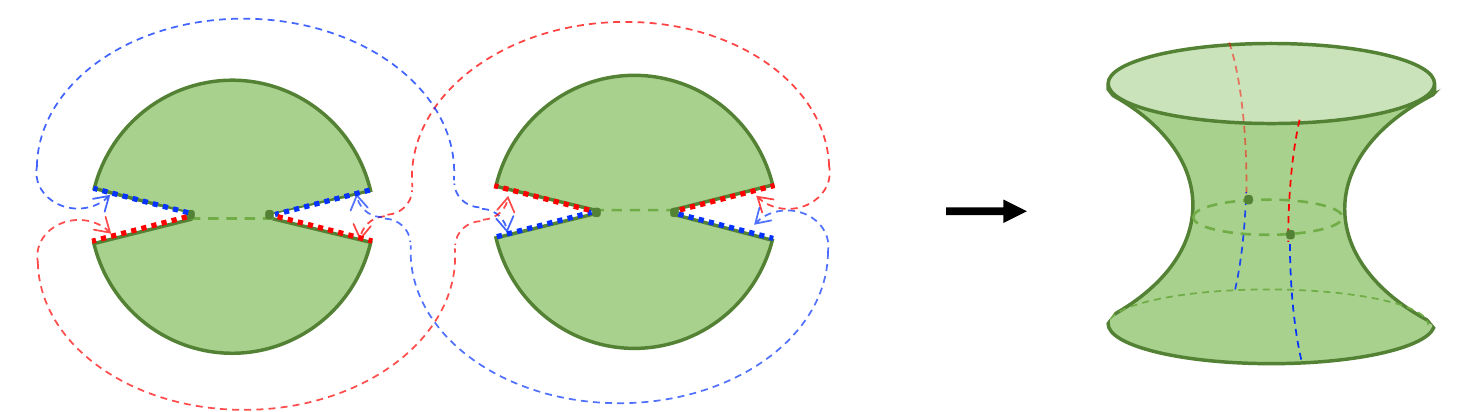}\nonumber \\[-4mm]\nonumber
\eea
as can be seen with a minor amount of mental exercise. The right-hand side exhibits two thermal circles, corresponding to the two replicas needed to compute the purity, connected through the bulk via a replica wormhole. The Island region is the half circle connecting to the dots at the equator of the wormhole. The two dots are the two quantum extremal surfaces that bound the Island region.

The same geometric exercise can be repeated for the higher Reny\'i entropies. For example, the saddle point geometry for the fourth Reny\'i entropy exhibits four replica wormholes all glued together into a single connected geometry \\[-4.5mm]
\bea
\tr(\rho_{\rm TMD}^4) \; \is\; \raisebox{-9mm}{\includegraphics[scale=0.475]{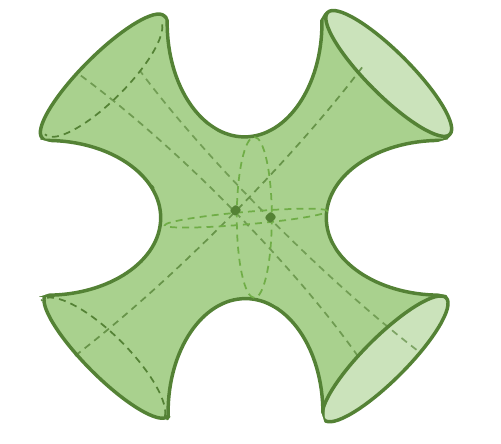}}\\[-3mm]\nonumber
\eea
The proposed identification \eqref{tmdsaddle} of the `janus pacman' geometry as the gravitational saddle point of the TMD state  \eqref{tmdt} predicts that the gravitational partition function of these connected multi-boundary geometries is given by the simple expression
\bea
\label{krenyi}
\tr(\rho_{\rm TMD}^{\spc k}) \is \sum_n p_n^k \spc = \spc  \frac{Z(k \beta)}{Z(\beta)^k}\, .
\eea
We now briefly describe how this prediction about the bulk gravity theory can be derived in the case of pure JT gravity or pure AdS3 gravity.

First consider the 3D case. For simplicity, let us assume that the holographic CFT lives on two circles and that the 3D bulk geometry describes a two sided BTZ black hole. The topology of the black hole horizon is therefore a single circle. 
Applying the above geometric prescription, the euclidean saddle point geometry for the $n$-th Reny\'i entropy has the following topology (for $n=4$)
\bea
\label{mthree}
{\cal M}_3 \; \is\; \raisebox{-9mm}{\includegraphics[scale=0.475]{island-renyi4x.pdf}} \ \times \ S^1. \\[-3mm]\nonumber
\eea
The boundary of this three-manifold ${\cal M}_3$ is the product of $n$ two-tori $T^2 = S^1 \times S^1$. One of these $S^1$'s is common to all $n$ boundary tori. The other $S^1$'s are the respective thermal circles, which are connected via the replica wormhole geometry, as shown. 

${\cal M}_3$ has the same euclidean BTZ black hole asymptotics near each boundary. The BTZ geometry is specified by a non-trivial space-time holonomy around the common $S^1$ direction. The holonomy takes values in the euclidean AdS3 isometry group $SL(2,\mathbb{C})$, and its conjugacy class is uniquely fixed by the mass and spin of the black hole. Moreover, the holonomy is a topological quantity: the conjugacy class of the holonomy group element is the same for any closed loop around the $S^1$. So the mass and angular momentum of all asymptotic BTZ geometries must all be identical. 

At the classical level, this identificiation does not seem at surprising or consequential, since all asymptotic regions are already identified via the replica symmetry.
In the quantum theory, the conjugacy class of the space-time holonomy around the $S^1$ is an operator. So eigenstates of the  holonomy operator correspond to energy and angular momentum eigenstates of the CFT.\footnote{This is not entirely accurate: the holonomy operators commute with the Virasoro generators, so when acting on a descendent state, they only measure the energy and angular momentum of the corresponding primary state. In AdS language, the holonomy operator is not sensitive to the presence of `boundary' gravitons. In the BTZ regime, however, almost all but an exponentially small fraction of the states are primary. So we are allowed to ignore this subtlety for our discussion here. For a more detailed treatment, see also the Appendix.} A classical black hole describes a canonical ensemble of mass and angular momentum eigenstates. A classical replica geometry as shown in equation \eqref{mthree} corresponds to a canonical ensemble of simultaneous mass and angular momentum eigenstates, weighted by the Boltzmann factor to the $n$-th power. Inside the ensemble, the mass and angular momentum is the  {\it same for all $n$ asymptotic boundaries} of the replica geometry, since they are all measured by the same holonomy operator.

The JT gravity case is  analogous and a special limit of the above. This is most directly seen by using the first order formulation of JT gravity developed in \cite{Iliesiu:2019xuh}. There it was shown that JT gravity can be exactly solved by writing it as a $BF$ gauge theory with gauge group given by a suitable central extension of  $PSL(2,\mathbb{R})$. The energy eigenstates of JT gravity are labeled by representations $R_s$ of $PSL(2,\mathbb{R})$, and the corresponding energy eigenvalues are proportional to the 2nd Casimir $C_2(R) = s^2-1/4$. Similarly as above, one can show that the energy operator at each boundary component descends from the same bulk operator $C_2 = \tr({B^2})$. Since the bulk geometry connects all boundary components, and the bulk $BF$-theory is topological, the energy eigenvalues of each boundary component must be identical to each other.\footnote{
Another perhaps more direct way to see that all replica sectors must carry the same quantum numbers is to consider the $PSL(2,\mathbb{R})$ holonomies $g_i$ around the $k$ $S^1$ boundaries of the replica wormhole space-time. To define these holonomies, we first cut open the wormhole geometry and represent it as a quotient of the Poincar\'e upper half plane. Since the sum of all $S^1$'s is a trivial cycle of the connected geometry, the JT functional integral will contain a delta function that implements the condition that the product $g_1 g_2 .. g_k$ of all holonomies must be trivial. This delta function can be written as an integral over $PSL(2,\mathbb{R})$ representations as
\bea
\label{delta}
\delta(g_1g_2 ... g_k-1) \is \int\! d\mu(s)\, \chi_s(g_1) \chi_s(g_2) ... \chi_s(g_k)
\eea
where $\chi_s(g) = \tr_{R_s}(g)$ defines the character of the representation $R_s$ and $d\mu(s)$ denotes the appropriate Plancherel measure on the space of $PSL(2,\mathbb{R})$  representations \cite{Iliesiu:2019xuh}.
The formula \eqref{delta} directly demonstrates that the connectedness of the bulk geometry implies that all replica sectors carry the same $PSL(2,\mathbb{R})$ representation and thus have the same energy. }

The final result for the JT gravity partition function  on a replica wormhole geometry with $k$ boundary components takes the same form  
 as in \eqref{krenyi} 
 \bea
 \label{jtrenyi}
\tr(\rho_{\rm JT}^{\, k}) \is \frac{Z(k\beta)}{Z(\beta)^k} \quad {\rm with} \quad Z(\beta) = \int\! d\mu(s) e^{-\beta s^2}
\eea
and $d\mu(s) = ds s \sinh(2\pi s)$ the spectral measure of JT gravity.
From  \eqref{jtrenyi} we immediately read off the that the density matrix of the two-sided black hole in JT gravity (in the post-Page time regime in which the Island prescription holds)  takes the TMD form
\bea
\label{jtrho}
\rho_{\rm JT} \is \frac{1}{Z(\beta)}  \int\! d\mu(s) \, e^{-\beta s^2} \li s \ra_L\la s\ri \otimes  \li s \ra_R\la s\ri, 
\eea
providing further support for our proposal.

\vspace{-3mm}
\subsection*{QEC, RT and Poincar\'e recurrence}
\vspace{-3mm}

Here we comment on the implications of our proposed holographic entropy relation for bulk reconstruction and its relation to quantum error correction.

It is useful to write the purification of the thermomixed double \eqref{tmdt} as an expansion in generalized TFD states  \eqref{tfdalpha} via $
|\Psi\ra  =  
\sum_\alpha \,|{\rm TFD}\ra_\alpha\, |\psi_\alpha\ra_{\cal E}$ 
with $|\psi_\alpha\ra_{\cal E}$ an orthogonal set of states of the ambient space-time ${\cal E}$. By tracing out ${\cal E}$, we recover the representation \eqref{tmdo} of the TMD state. The above expansion of $|\Psi\ra$ can be rewritten in terms of time-evolved TFD states 
\bea
|\Psi\ra\, \is\,  
\sum_{t_\alpha} \, |{\rm TFD}(t_\alpha)\ra \, |t_\alpha\ra_{{}_{\cal E}}
\eea 
with $t_\alpha$ the time variable associated with $\alpha$ via  \eqref{talpha}. We abbreviate the  state of the environment ${\cal E}$ by $|t_\alpha\ra_{{}_{\cal E}}$, to indicate that for our purpose, ${\cal E}$  can simply be thought of as a clock that keeps track of the age of the two sided black hole. All state $|t_\alpha\ra$ are assumed to be orthogonal, so the clock has a very large Hilbert space. We introduce a time evolution operator $U(t)$ such that
\bea
\label{timev}
|{\rm TFD}(t)\ra \is \, U(t)\spc |{\rm TFD}\ra\nonumber\\[-1mm]\\[-1mm]\nonumber
U(t)\spc =& & \!\!\!\!\!\!    e^{-it (H_L + H_R)}
\eea
with $H_L$ and $H_R$ the left and right Hamiltonian.
The variable $t$ runs over a range equal to the Poincar\'e recurrence time $\tau_P^{\rm cft}$ of the CFT dual to two-sided black hole. This time is double exponentially large in the entropy $S_{BH}$. 

Two observers, Alice and Bob, want to measure the age $t$ of the two sided black hole. As explained above, they can do this by first synchronizing their clocks and jumping in. Suppose that they are lucky enough to meet inside the black hole. They can then compare their clocks and thereby measure the age of the two-sided black hole.

As we will now argue, the above story is not fully~correct. 
Alice and Bob are part of the low energy effective QFT of the near horizon region. The QFT Hilbert space spans a small code subspace ${\cal H}_{\rm code}$ inside the full microscopic CFT Hilbert space. This code subspace is much smaller than that of the CFT, and also much smaller than the clock Hilbert space spanned by all the states $|t_\alpha\rangle$. How can a smaller system make an accurate measurement of the state a much larger system? 

As mentioned above, the range of time $t$ extends to the microscopic Poincar\'e recurrence time $\tau_P^{\rm cft}$. The low energy QFT, however, can record time $t$ at most up to its own Poincar\'e recurrence time $t_P^{\rm qft}$. Evidently, the latter is much smaller than the former
\bea
\log t_P^{\rm qft} \simeq {e^{S_{\rm code}}} \; \ll \;\, \log t_P^{\rm cft} \simeq {e^{S_{BH}}}
\eea
Because of this hierarchy in time scales, there are many time evolution operators $U(t_{\bar\alpha})$ that act as identity operator on the code subspace,  and thus commute with all low energy effective QFT operators
\bea
\label{comrel}
[U({t_{\overline{\alpha\nspc}}}), {\cal O}_{\rm qft}] \is 0, \qquad \mbox{within ${\cal H}_{\rm code}$}.
\eea
Hence we conclude that  Alice and Bob can at best measure the age $t$ of the two-sided black hole only up to multiples of the Poincar\'e recurrence time of the low energy QFT. This still leaves a large uncertainty in the value of~$t$.

We can make this claim a bit more explicit as follows. Let $\hat{K} = - \log \rho_R$ denote the modular Hamiltonian of the right CFT. When restricted to act within the code subspace, this microscopic modular Hamiltonian reduces to the modular Hamiltonian of the low energy QFT inside the right bulk region, up to a constant term equal to the area of the horizon \cite{FLM}\cite{DHW}
\bea
P_{\rm code} \hat K P_{\rm code} \is \frac{A}{4 G_N} P_{\rm code}  +  \hat{K}_{\rm qft} 
\eea
This relation implies that the microscopic time evolution operator $U(\tau) \spc =\spc e^{-2i \tau \hat{K}}$ and the effective QFT time evolution operator $ U_{\rm qft}(\tau) \spc = \spc  e^{-2i\tau \hat{K}_{\rm qft}}$
are identical, up to an overall phase factor, when restricted to the code subspace
\bea
\label{projectrel}
 U_{\rm qft}(\tau) \is P_{\rm code} U(\tau) P_{\rm code}.
\eea
Now consider the relation $U_{\rm qft}( \tau^{\rm qft}_P) = P_{\rm code}$ that defines the Poincar\'e recurrence time of the low energy QFT. Using \eqref{projectrel}, we deduce that for all integer $n$
\bea
\label{period}
U (n \tau^{\rm qft}_P) \is \mathbb{1} \qquad \mbox{within ${\cal H}_{\rm code}$}.
\eea
So we confirm that \eqref{comrel} holds for all multiples $t_{\overline\alpha} = n \tau^{\rm qft}_P$ of the Poincar\'e recurrence time of the low energy QFT.

Let $|{\rm tfd} \ra$ denote the thermofield double state (Unruh vacuum) of the low energy effective QFT. We  like to construct the balanced mixed state  with the largest entropy  that, when restricted to the code subspace, reduces to this tfd-state.

Consider the mixed state given by the incoherent sum of special time evolved TFD states
\bea
\rho_{\,\overline{\rm\! TMD\!}} \, \is \frac 1 N \sum_{t_{\overline{\alpha\nspc}}}  |{\rm TFD}(t_{\spc\overline{\nspc\alpha\nspc}})\ra\la {\rm TFD}(t_{\spc\overline{\nspc\alpha\nspc}})|
\eea
with ${t_{\overline{\alpha\nspc}}} = n \tau^{\rm qft}_P$. Since there are many such times,  the above state looks, up to very good approximation, like the TMD state \eqref{tmdt}. In particular, it has von Neumann entropy $S_{BH}$ and exhibits mostly only classical correlations. However, from equations \eqref{timev} and \eqref{period}, we deduce that this state has the desired property that, when restricted to the code subspace, it reduces to the pure tfd-state of the effective QFT
 \bea
P_{\rm code} \, \rho_{\,\overline{\rm\! TMD\!}} \; P_{\rm code}  \is |{\rm tfd}\ra \la {\rm tfd}|. 
\eea
 So $\rho_{\,\overline{\rm\! TMD\!}}$ still encodes coherent quantum entanglement, but only within the code subspace. This is sufficient to ensure the existence of the ER bridge.

A similar class of operators with property \eqref{comrel} were introduced in  \cite{preskill}, where they  are called `ghost operators'.

\vspace{-3mm}
\subsection*{Decoherence of a two-sided black hole}
\vspace{-3mm}

Our proposal makes  a concrete prediction about the time evolution and decoherence of an evaporating two-sided black hole due to the Hawking process. 
Denote the Hilbert space of the radiation by ${\cal H}_{{}_{\mathsf{{\cal E}}}}$.  A priori, a generic entangled state of the black hole plus environment can take the general form
\bea
|\Psi\ra \is \sum_{n,\tilde{n}} \, |n,\tilde{n}\rangle_{{}_{\!\spc\mathsf{CFT2}}} \, |\phi_{n\tilde{n}}\rangle_{{}_{\!\spc{{\cal E}}}}
\eea
with $|n,\tilde{n}\rangle_{{}_{\!\spc\mathsf{CFT2}}}  = |n\ra_L |\tilde{n}\ra_R$ the state of the two CFTs, and where $|\phi_{n\tilde{n}}\rangle_{{}_{\!\spc{{\cal E}}}}$ denotes the state of the environment. The corresponding density matrices of the pair of CFTs reads
\bea
\rho_{{}_{\mathsf{CFT2}}}\spc =\! \sum_{n,\tilde{n},m,\tilde{m}}  
\, \langle\phi_{n\tilde{n}}|\phi_{m\tilde{m}}\rangle\;  |n,\tilde{n}\rangle \langle m,\tilde{m}|
\eea
It is a reasonable physical assumption that the internal dynamics of a two-sided black hole is ergodic, possibly within a physical subspace, and that the matrix elements of the time evolution operator constitute some random matrices, taken from some suitable matrix ensemble. 

We contrast two special choices of the ensemble average of the matrix element 
\bea
\ \ \overline{\!\strut\langle\phi_{n\tilde{n}}|\phi_{m\tilde{m}}\rangle\!}
\, \is \left\{\begin{array}{c}
\quad\;\, {p_{n}\, p_{\tilde{n}}\, \delta_{nm}\, \delta_{\tilde{n}\tilde{m}} \quad \ \ \ \ (\rm{case\ 1})}\\[2.5mm]
\quad\;\, {p_{n} \, \delta_{n\tilde{n}}\, \delta_{nm}\, \delta_{n\tilde{m}} \quad  \ \ \;(\rm{case\ 2})}\end{array}\right. \nonumber
\eea
The corresponding two-sided CFT density matrices are
\bea 
\label{cases}
\ \rho_{{}_{\mathsf{CFT2}}}
 \is \left\{\begin{array}{c} 
\    \quad \, \rho_{L} \otimes \, \rho_{R} \quad  \quad  \ \  (\rm{case\ 1})\\[2.2mm]
 \qquad \ \, \rho_{{\rm TMD}}\; \quad \quad  \ \    \spc (\rm{case\ 2})
\end{array}\right.
\eea 
In case 1 each CFT independently decoheres, by building up entanglement with its emitted Hawking radiation. After the Page time, each is in a mixed thermal state. The total von Neumann entropy of the two sided black hole in this case is $2S_{BH}$. Note, however, that the two CFTs no longer share is any mutual information. 

In case 2 each CFT is also in a thermal state, but shares half of its entanglement with the radiation and half of its entanglement with the other CFT. The resulting thermo-mixed double state  describes the universal mixed state of a two sided black hole that arises after full dephasing \cite{tadashitmd}, while maintaining the `balanced' condition that both CFTs have the exact same energy \cite{balanced}.

What mechanism could produce this `balanced' type of decoherence? In \cite{balanced}, the idea was put forward that black hole information is stored non-locally, similar to a topological quantum memory like the toric code. This analogy is most apparent in the above description of generalized TFD states. With this perspective in mind, let us briefly indicate how a `balanced' decoherence could arise.

Suppose that until time $t=0$, the two CFTs are decoupled from the environment ${\cal E}$. At $t=0$, we turn on the interaction. Assuming ${\cal E}$ starts in its ground state $|0\ra_{{}_{\!\spc{{\cal E}}}}$, the time evolved state of some arbitrary initial CFT state $|i,\tilde{i}\spc\ra$ takes the schematic form
\bea
\label{step}
U |i,\tilde{i}\spc\ra_{{\!\spc}_{\mathsf{CFT2}}} |0\ra_{{}_{\!\spc{{\cal E}}}}\! \is\! \sum_{k,\tilde{k},r,\tilde{r}} \spc C^{i}_{kr}C^{\tilde{i}}_{\tilde{k}\tilde{r}}|k,\tilde{k}\ra_{{\!\spc}_{\mathsf{CFT2}}} \, |r,\tilde{r}\ra_{{}_{\!\spc{{\cal E}}}}
\eea
The coefficients $C^{i}_{kr}$ can be thought of as structure coefficients of an operator algebra (OPE). A priori, they could be fully random. However, suppose that (just like OPE coefficients) they  generate some associative and commutative algebra. Similar to the Verlinde formula, the  $C^{i}_{kr}$ matrices are then simultaneously diagonalizable via some unitary rotations $S$ and $V$ applied to the CFT and environment ${\cal E}$, resp:
\bea
C^i_{kr} \is \sum_n \lambda_n\spc S^{*i}_{\spc n}\spc S_k^n  V_r^n
\eea
with $\lambda_n$ some set of eigenvalues. After going to the diagonal basis $|n,\tilde{n}\ra_{{\!\spc}_{\mathsf{CFT2}}} = \sum_{i,\tilde{i}} S_i^{n} S^{\tilde{n}}_{\tilde{i}}\spc |i,\tilde{i}\ra_{{\!\spc}_{\mathsf{CFT2}}}$ on the CFT and $|\phi_{n\tilde{n}}\ra_{{}_{\!\spc{{\cal E}}}} = \lambda_n \tilde{\lambda}_{\tilde{n}} \sum_{r,\tilde{r}} V_r^{n}V_{\tilde{r}}^{\tilde{n}}\spc |r,\tilde{r}\ra_{{}_{\!\spc{{\cal E}}}}$ on the environment, the time evolution step \eqref{step} diagonalizes. So starting with the TFD state, 
this time evolution will result in the case 2 scenario of decoherence:
\bea
U |n,{n}\ra_{{\!\spc}_{\mathsf{CFT2}}} |0\ra_{{}_{\!\spc{{\cal E}}}}\spc \is\spc 
|n, n\ra_{{\!\spc}_{\mathsf{CFT2}}} |\phi_{nn}\ra_{{}_{\!\spc{{\cal E}}}}
\eea
 The above diagonalization mechanism is formally similar to Coleman's introduction of the $\alpha$ eigenstates of the wormhole creation operators \cite{Coleman}\cite{DonHenry}.

What if instead we set up a dynamical time evolution such that the two CFTs completely decohere into the factorized thermal state? The $n$-th Reny\'i entropy of the final state then also factorizes: $\tr\bigl(( \rho_{L} \otimes \, \rho_{R})^n\bigr) = \tr( \rho_{L}^{\spc n}) \; \tr( \, \rho^{\spc n}_{R}).$
The corresponding saddle point geometries describe the disconnected $n$-th replica manifolds of two disconnected single sided black holes. The ER bridge has disappeared: the typical state in the factorized thermal ensemble has a firewall \cite{FW} that splits the two-sided black hole into to two separate one-sides black hole geometries. 
Indeed, the firewall problem for the two sided black hole comes back in full force if the ensemble of accessible quantum states includes states in which the left- and right states have different energy. Such states can not be written as linear combinations of generalized TFD states. 

\vspace{-3mm}
\subsection*{Conclusion}
\vspace{-3mm}

We have formulated a new holographic identification between the cross-sectional area of an Einstein-Rosen bridge of a two-sided black hole and the maximal  amount of quantum information contained inside the black hole. We have given evidence in support of this proposal in the context of low dimensional gravity models, and via a general argument based on the relationship between bulk reconstruction and quantum error correcting codes.

According to our proposal, the connectedness of space-time across an ER-bridge does not require that the black hole region is in a pure maximally entangled TFD state. Instead, the region around the ER bridge can be in a thermal mixed double state with $S_{BH}$ worth of classical correlations between the two sides; it is sufficient that the quantum entanglement only resides in the code subspace containing the low energy effective QFT.  

The balanced holography principle implies that the quantum information inside the black hole and inside the Island region is topologically protected \cite{balanced}. This protection is dynamically implemented via a restriction that prevents the black hole from accumulating more entropy than $S_{BH}$. Instead the total state will evolve into a GHZ like state of the form \eqref{psi}. In such a state, the quantum information stored inside the Island region can not be accessed from any one of the three subregions $L$, $R$ or ${\cal E}$. Instead, it is  accessible only through combined knowledge of the state of at least two of the three subregions. Hence the reconstruction of the QFT operators inside the Island region  involves a similar quantum error correction mechanism as for bulk reconstruction in AdS/CFT \cite{EWR}.

\begin{figure}[t]
\begin{center}
\vspace{-2mm}
\begin{tikzpicture}[scale=.88]
\draw [black!0,fill=yellow!025,dashed,thin,domain=0:360] plot ({2* cos(\x)},{2 * sin(\x)});
\draw [black!0,fill=purple!35!yellow!20!red!25,dashed,thin,domain=-30:90] plot ({2* cos(\x)},{2 * sin(\x)});
\draw [black!0,fill=purple!35!yellow!20!red!25,dashed,thin,domain=90:210] plot ({2* cos(\x)},{2 * sin(\x)});
\draw [black!0,fill=yellow!025,dashed,thin,domain=210:330] plot ({2* cos(\x)},{2 * sin(\x)});
\draw [black,fill=purple!35!yellow!20!red!25,dashed,thick,domain=210:150] plot ({cos(-30) * (4  + 3.46 * cos(\x) ) + sin(-30) * (3.46 * sin(\x)  )}, {-sin(-30) * (4  + 3.46 * cos(\x) ) + cos(-30) * (3.46  * sin(\x)  )});
\draw [red,fill=yellow!025,dashed,thick,domain=210:150] plot ({cos(90) * (4  + 3.46 * cos(\x) ) + sin(90) * (3.46  * sin(\x)  )}, {-sin(90) * (4  + 3.46 * cos(\x) ) + cos(90) * (3.46  * sin(\x)  )});
\draw [black,fill=purple!35!yellow!20!red!25,dashed,thick,domain=210:150] plot ({cos(210) * (4  + 3.46 * cos(\x) ) + sin(210) * (3.46  * sin(\x)  )}, {-sin(210) * (4  + 3.46 * cos(\x) ) + cos(210) * (3.46  * sin(\x)  )});
\draw [black,thin] (0,0) circle (2cm);
\draw [fill=green!80,color=green!80] (-1,.63) circle (1.3mm);
\draw [fill=green!80,color=green!80] (-1,.3) circle (2mm);
\draw [fill=green!80,color=green!80] (-1.2,.1) -- (-1.2,.3) -- (-.8,.3) -- (-.8,.1);
\draw (-1,.3) node {$\mbox{\small \bf  A}$};
\draw [fill=red!60,color=green!52] (1,.63) circle (1.3mm);
\draw [fill=red!60,color=green!52] (1,.3) circle (2mm);
\draw [fill=red!60,color=green!52] (.8,.1) -- (.8,.3) -- (1.2,.3) -- (1.2,.1);
\draw (1,.3) node {{$\mbox{\small \bf  B}$}};
\draw[->,thick][color={black}](-.7, .3) -- (-.2, .2);
\draw[->,thick][color={black}](.7,.3) -- (.2, .2);
\draw (-1.2,1.15) node {$\mbox{\large $L$}$};
\draw (1.2,1.15) node {$\mbox{\large $R$}$};
\draw (0,-1.3) node {$\mbox{\large ${\cal E}$}$};
\draw (0,-.2) node {$\mbox{\footnotesize \bf  Island}$};
\end{tikzpicture}
\end{center}
\vspace{-2mm}
\caption{The Island is contained inside the entanglement wedge of any pair of the three regions $L$, $R$ and ${\cal E}$. The two regions $L$ and $R$ combined are in the thermal mixed state, with entropy given by the area of the red RT surface.}
\vspace{-3mm}
\end{figure}
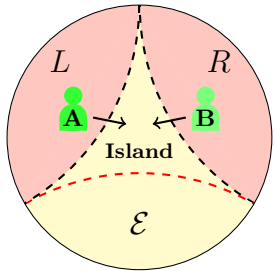

Our new holographic principle stands at odds with the strong version of ER~=~EPR, which equates the existence of a smooth ER-bridge with the presence of microscopic entanglement, quantified by the Berkenstein-Hawking-Ryu-Takayanagi entropy formula \eqref{bh}. So what gives? It is important to point out, however, that the strong ER~=~EPR implication amounts to a radical departure from the usual rules of quantum mechanics: entanglement is not a linear property of quantum states, so it is highly unconventional to associate it with an observable property of a macroscopic system. The aim and spirit of our proposal is to refine the ER~=~EPR relation so that it does not run counter to any conventional ground rules of quantum mechanics.

The fact that the near-horizon region of a two sided black hole can be in a mixed state plays a key role in recent discussions of the Island rule for computing the entropy  of an evaporating black hole and its radiation. Motivated by the new observation that an evaporating two-sided black hole geometry supports a two-component quantum extremal surface demarcating an Island region inside the interior, the authors of \cite{replica2} propose the following generalization of the holographic entropy formula for the von Neumann entropy of the near-horizon region of a two sided black hole
\bea
\label{islandone}
S(L\cup R)\, \is 
{\rm ext}_Q \left[ \frac{{\rm Area}(Q)}{4G_N}  \; + \; S_{\rm qft}(L\cup R)\right]\ \ 
\eea
Here $Q$ indicates the location of the quantum extremal surface \cite{Penington:2019npb}-\cite{Almheiri:2019psf}.  By definition, this location extremizes the sum of the area contribution of the quantum extremal surface and the von Neumann entropy of the effective QFT state inside the near-horizon region $L\cup R$ \cite{EngelhardtWall}.

\begin{figure}[t]
\begin{center}
\vspace{-.1cm}
${}$\ \  \includegraphics[scale=0.37]{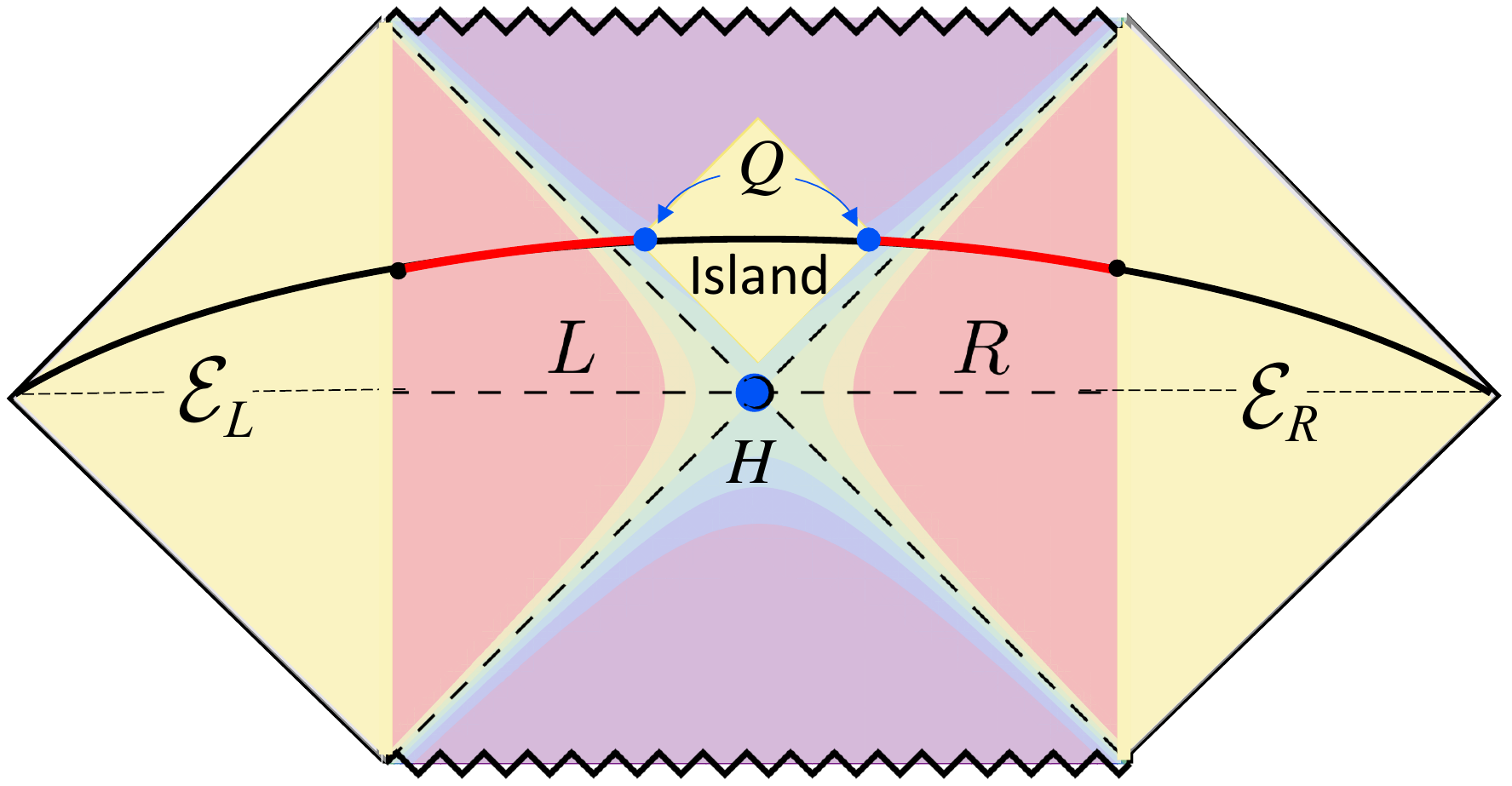}
\end{center}
\vspace{-1mm}
\caption{Penrose diagram of an ER bridge with an Island region, bounded by two quantum extremal surfaces.}
\vspace{-2mm}
\end{figure}

As noted in \cite{replica2}, equation  \eqref{islandone} has a complement in the form of the following formula for the von Neumann entropy of (the Hawking radiation contained inside) the environment ${\cal E}$
\bea
\label{islandtwo}
S({{\cal E}})\, \is 
{\rm ext}_Q \left[ \frac{{\rm Area}(Q)}{4G_N}  \; + \; S_{\rm qft}(I \cup {\cal E})\right]\ \ 
\eea
The right-hand side of \eqref{islandone} and \eqref{islandtwo} are both identical -- as it should, since the total state on $L\cup R \cup {\cal E}$ is assumed to be a pure state. 

How do equations \eqref{islandone} and \eqref{islandtwo} stand in relation to with our proposed holographic bound? The quantum extremal surface $Q$ jumps into existence right at the Page time, i.e. at the future time slice at which the entropy of the black hole becomes equal or smaller than the entropy of the emitted Hawking radiation. Assuming the naive Page dynamics applies, this transition happens at the moment the black hole has transferred half its initial entropy to the radiation, or in other words, when the black hole horizon area has decreased to half the initial size. Based on this physical reasoning, we would infer that the two components of the quantum extremal surface $Q$ combined have the same total area as the event horizon $H$ of the initial two-sided black hole at $t=0$
\bea
\frac{{\rm Area}(Q)}{4G_N} \, \simeq\,\frac{{\rm Area}(H)}{4G_N} \spc \is \spc S_{BH}
\eea
This relation would align the analysis of \cite{replica2} with the present story: the left-hand side equals the entropy of the near-horizon region of the two-sided black hole, and the equality states that in the stationary post-Page time situation, it saturates our new holographic bound.\footnote{Note, however, that the description of the entanglement and decoherence generated through the Hawking process presented in \cite{replica2} is still different from ours. In the scenario outlined in \cite{replica2}, the left- and right CFT each independently decohere, and evolve into the factorized thermal state with von Neumann entropy $2S_{BH}$. This is the case 1 scenario introduced above. We interpret this case 1 scenario as giving rise to two disconnected black hole geometries without an ER bridge.\\}

In our description, the state of the old two sided black hole settles down in a stationary TMD state. Hence there is no notion of holographic complexity that keeps growing in time\footnote{Let us compare this with the idealized AdS/CFT time evolution in terms of time evolved or generalized TFD states $\li  \mbox{\small \sc TFD} \ra\raisebox{-1.5pt}{${\nspc}_{\alpha}$}$. In this pure state description, one could introduce a notion that behaves similar to complexity, by taking minus the logarithm of the spectral form factor
$$
Z[\alpha] = \li \la   \mbox{\small \sc TFD} \li  \mbox{\small \sc TFD} \ra\raisebox{-1.5pt}{${\nspc}_{\alpha}$}\ri^2\! = 
\sum_{m,n} p_n p_m e^{i (\alpha_n\!\smpc-\alpha_m)}.\ \ 
$$
Setting $e^{i\alpha_n} = e^{-2it E_n}$,  this quantity would initially grow linearly with time, just like the volume of the interior. In the final incoherent ensemble of generalized TFD states,  the averaged value of the spectral form factor reduces to the overlap of the TFD and TMD state
$$
\overline{Z[\alpha]} = \frac{1}{N}\sum_{\alpha}  Z[\alpha] 
= \tr(\rho_{\rm TFD} \rho_{\rm TMD}) \spc = \spc \frac{Z(2\beta)}{Z(\beta)^2}
$$
which equals the purity of the TMD state. This is the plateau value of the spectral form factor \cite{formfactor} and the final stationary value of this notion of complexity.}. From the perspective presented here, the notion of holographic complexity put forward in \cite{complexity} relies on an idealized identification of classical space-time regions with pure states.

\vspace{-3mm}

\section*{Acknowledgements}
\vspace{-3mm}

I thank  Ahmed Almheiri, Vijay Balasubramanian, Netta Engelhardt, Daniel Harlow, Juan Maldacena, Rob Myers, Jonathan Oppenheim, Xiaoliang Qi, Douglas Stanford, Tadashi Takayanagi and Erik Verlinde for useful discussions and comments. This research is supported by NSF grant PHY-1620059.

\vspace{-1mm}
\subsection*{Appendix: Edge states and topological entanglement}
\vspace{-2mm}

The idea that black hole information is topologically protected receives further support from the observations that the formula \cite{KP}  for topological entropy  for Virasoro CFTs reproduces the Bekenstein-Hawking entropy \cite{LH}, and the insight that entanglement wedge reconstruction involves mechanisms akin to quantum secret sharing \cite{EWR}\cite{VJ},  and that the bulk state in AdS/CFT or FQHE theories can be modeled via tensor networks \cite{Happy} or matrix product states \cite{zaletel}.

In this Appendix, we further substantiate our argument in the main text by generalizing results from the study of topological entanglement entropy in FQHE systems and Chern-Simons field theory \cite{QKL}\cite{shinsei}. In CS theories, the Hilbert space does not simply factorize. This obstruction is encoded in the form of universal long range entanglement, that only depends on the topology of the space time \cite{KP}. Moreover, on a manifold with a boundary, CS theory gives rise to massless edge degrees of freedom. 3D AdS gravity shares the same formal properties. The massless edge degrees of freedom in this case constitute the holographic dual (Virasoro) CFT.

\begin{figure}[t]
\begin{center}
\vspace{0cm}
\includegraphics[scale=1.15]{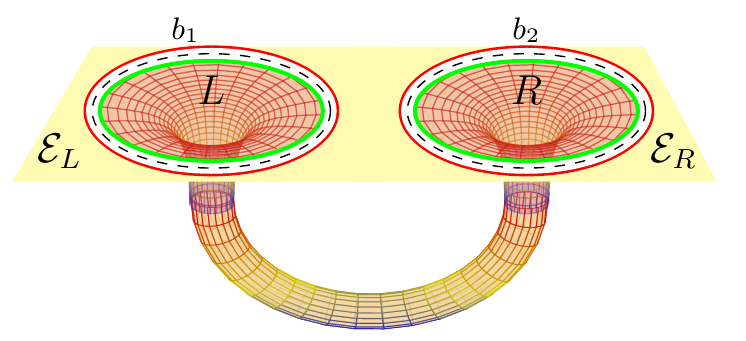}
\end{center}
\vspace{-1mm}
\caption{The two entanglement cuts surrounding the near horizon region of the two-sided black hole introduce edge states on each side of the cut. The combined state of edges  across each cut are described by a boundary state of the holographic CFT.}
\vspace{-2mm}
\end{figure}

Suppose we introduce an entanglement boundary by cutting space-time along two circles surrounding the left and right near horizon regions, as depicted in Fig. 5. The physical step of making the entanglement cut amounts to performing a quantum quench in which the interaction Hamiltonian, that connects the two sides of the cut, is suddenly switched off \cite{QKL}. The remaining Hamiltonian then splits up into two commuting terms, each acting only on one side of the cut. 
Turning off the interaction across the cut liberates the edge dynamics of the two bulk regions and introduces massless modes running along each side of the cut \cite{KP}\cite{QKL}.   In CS theory, these massless excitations are chiral. In AdS3/CFT2, the massless modes on the interior edges are identified with the holographic CFT dual to the black hole region, while the massless modes on the exterior edges are holographic dual modes associated with the ambient space-time. Both holographic edge theories are non-chiral.  

As argued in \cite{QKL}\cite{shinsei}, following work by Cardy and Calabrese \cite{cardy-calabrese}, the state across an entanglement cut takes the form of a boundary CFT state that pairs up the chiral edge modes (indicated with green and red circles in Fig. 5) on both sides of the cut. For CS theory, there's only one boundary state for each cut; for 3D AdS gravity, each cut is represented by the tensor product of two conjugate CFT boundary states. To simplify the notation, we write the formulas only for the case of the CS theory.

In CS theory, the combined state of the two entanglement cuts takes the following form \cite{shinsei}
\bea
\label{ishidouble}
\li \Psi \ra \is \sum_n \psi_n | \mathfrak{h}^{1}_n\ra\!\ra \otimes | \mathfrak{h}^{2}_n\ra\! \ra \nonumber\\[-2.5mm]\\[-2.5mm]\nonumber
& & \hspace{-7mm} | \mathfrak{h}^{i}_n\ra\!\ra \spc = \spc  \frac{e^{- {\epsilon} H_i}} {\sqrt{\mbox{\small $\raisebox{-1pt}{\normalsize $Z$}_{h^{i}_n}\!(\epsilon)$}\!}}\,  | {h}^{i}_n\ra\!\ra
\eea
with $| {h}^{i}_n\ra\!\ra$ a complete basis of Ishibashi boundary states~at each entanglement cut $b_i$, $\psi_n$ with $
\sum_n |\psi_n|^2 = 1 
$ a normalized set of amplitudes, 
$H_i\! = L_0^i\!+ \overline{L}_0{\!\!}^i-\frac{c}{12}$\, the respective~CFT Hamiltonian,  $\epsilon$ a UV regulator, and $\mbox{$Z_{h^i_n}\!(\epsilon)$}=\la\!\la {h}^{i}_n|  e^{-\epsilon H_i}| {h}^{i}_n\ra\!\ra$.

Ishibashi states are entangled sums of left- and right-moving descendent states $|N,h\rangle \otimes |N,\tilde{h}\rangle$ of primary states $|h\rangle |\tilde{h}\ra$ with the same left- and right-moving conformal dimension $h=\tilde{h}$. Hence after tracing out the exterior edge degrees of freedom associated with the ambient space-time ${\cal E}$, we obtain a reduced density matrix of the thermo-mixed double form \cite{shinsei}
\bea
& & \ \ \rho  \, = \, \sum_n \, |\psi_n|^2 \rho_n^1 \otimes \rho_n^2\\[.5mm]
\rho_n^{i} \is\frac 1 {Z_{h^{i}_n}\!(\epsilon)\!}\, \sum_{N} e^{-2\epsilon H_i} |N,h_n^{i}\rangle \langle h_n^{i} ,N|
\eea
From this explicit form of the density matrix, it is straightforward to derive the following result for the $k$-th Renyi entropy \cite{shinsei}
\bea
\label{renyit}
\tr(\rho^k) \is e^{\mbox{\small $\frac{\pi^2\!\spc c}{1\! \spc 2\epsilon}$}(k-\frac{1}{k})} \sum_n |\psi_n|^{2k} |S_{0}^n|^{2-2k}
\eea
where $S_0^n$ denotes the matrix element of the modular\ $S$-matrix of the CFT, and we used that the leading contribution for $\epsilon\to 0$ comes from the vacuum sector in the dual channel.
The divergent prefactor is the non-universal UV contribution from short range entangled modes across both entanglement cuts. The  sum over $n$ is the topological contribution to the Reny\'i entropy.

For holographic CFTs, the sum over $n$ is dominated by the dense spectrum of black hole states. As shown by the above derivation, the density of states is proportional to the square of the modular S-matrix element $S_0^n$. For Virasoro CFTs the matrix element $S_0^n$ is equal to the Cardy-Bekenstein-Hawking density of states $e^{S_{BH}}$ \cite{LH}. In this regime, it is appropriate to rewrite the above expression for the Reny\'i entropy by introducing the normalized probability distribution  
and spectral density via 
\bea
p_n =\spc {|\psi_n|^2}/{|S_0^n|^2} \quad & & \quad
\int\! d\mu(n) \, ...\,  = \, \sum_n |S_0^n|^2 .... \ \ \nonumber
\eea
Making this substitution in \eqref{renyit} and dropping the cut-off dependent prefactor give 
\bea
\tr(\rho^{\spc k}) \is  
\int\! d\mu(n) \, p_n^{\, k}.
\eea
We thus again reproduce the expression \eqref{krenyi} for the $k$-th Renyi entropy of the TMD state.


\begin{thebibliography}{99}

\bibitem{bh} S. Hawking, ~~Particle Creation by Black Holes", 
Commun.Math.Phys. 43 (1975) 199;   J.~D.~Bekenstein, ``Black holes and entropy,'' 
 Phys.\ Rev.\ D {\bf 7}, 2333 (1973).

\bibitem{RT} 
  S.~Ryu and T.~Takayanagi,
  ``Holographic derivation of entanglement entropy from AdS/CFT,''
  Phys.\ Rev.\ Lett.\  {\bf 96}, 181602 (2006)
  
  
  
  \bibitem{Maldacena:2001kr} 
  J.~M.~Maldacena,
  ``Eternal black holes in anti-de Sitter,''
  JHEP {\bf 0304}, 021 (2003)
  
 \bibitem{vanraamsdonk}  M. Van Raamsdonk, Comments on quantum gravity and entanglement, arXiv:0907.2939

  \bibitem{EH}{E. Verlinde and H. Verlinde, unpublished}
  
\bibitem{ER=EPR} 
  J.~Maldacena and L.~Susskind,
  ``Cool horizons for entangled black holes,''
  Fortsch.\ Phys.\  {\bf 61}, 781 (2013)
  
  


\bibitem{Penington:2019npb} 
  G.~Penington,
  ``Entanglement Wedge Reconstruction and the Information Paradox,''
  arXiv:1905.08255 .
  
\bibitem{Almheiri:2019psf} 
  A.~Almheiri, N.~Engelhardt, D.~Marolf and H.~Maxfield,
  ``The entropy of bulk quantum fields and the entanglement wedge of an evaporating black hole,''
  arXiv:1905.08762 .
  

  \bibitem{Almheiri:2019hni} 
  A.~Almheiri, R.~Mahajan, J.~Maldacena and Y.~Zhao,
  ``The Page curve of Hawking radiation from semiclassical geometry,''
  arXiv:1908.10996 .
  
  
\bibitem{Akers:2019nfi} 
  C.~Akers, N.~Engelhardt and D.~Harlow,
  ``Simple holographic models of black hole evaporation,''
  arXiv:1910.00972 .


\bibitem{Almheiri:2019yqk} 
  A.~Almheiri, R.~Mahajan and J.~Maldacena,
  ``Islands outside the horizon,''
  arXiv:1910.11077 .
  
  
\bibitem{replica1} 
  G.~Penington, S.~H.~Shenker, D.~Stanford and Z.~Yang,
  ``Replica wormholes and the black hole interior,''
  arXiv:1911.11977 .
  
 
\bibitem{replica2} 
  A.~Almheiri, T.~Hartman, J.~Maldacena, E.~Shaghoulian and A.~Tajdini,
  ``Replica Wormholes and the Entropy of Hawking Radiation,''
  arXiv:1911.12333 .
  


\bibitem{VV-QEC} 
  E.~Verlinde and H.~Verlinde,
  ``Black Hole Entanglement and Quantum Error Correction,''
  JHEP {\bf 1310}, 107 (2013),
  arXiv:1211.6913.
  
\bibitem{balanced} 
  E.~Verlinde and H.~Verlinde,
  ``Passing through the Firewall,''
  arXiv:1306.0515 ;
  E.~Verlinde and H.~Verlinde,
  ``Black Hole Information as Topological Qubits,''
  arXiv:1306.0516 .

\bibitem{EWR} 
  A.~Almheiri, X.~Dong and D.~Harlow,
  ``Bulk Locality and Quantum Error Correction in AdS/CFT,''
  JHEP {\bf 1504}, 163 (2015)
  
  
\bibitem{Beni} 
  B.~Yoshida,
  ``Firewalls vs. Scrambling,''
  JHEP {\bf 1910}, 132 (2019); ``Observer-dependent black hole interior from operator collision,''
  
\bibitem{Ahmed} 
  A.~Almheiri,
  ``Holographic Quantum Error Correction and the Projected Black Hole Interior,''
  arXiv:1810.02055.


\bibitem{VJ} 
  V.~Balasubramanian, A.~Kar, O.~Parrikar, G.~Sárosi and T.~Ugajin,
  ``Geometric secret sharing in a model of Hawking radiation,''
  arXiv:2003.05448.


\bibitem{Iliesiu:2019xuh} 
  L.~V.~Iliesiu, S.~S.~Pufu, H.~Verlinde and Y.~Wang,
  ``An exact quantization of Jackiw-Teitelboim gravity,''
  JHEP {\bf 1911}, 091 (2019)
  
    \bibitem{preskill} 
  I.~H.~Kim, E.~Tang and J.~Preskill,
  ``The ghost in the radiation: Robust encodings of the black hole interior,''
  arXiv:2003.05451.
  
  
  \bibitem{FLM} 
  T.~Faulkner, A.~Lewkowycz and J.~Maldacena,
  ``Quantum corrections to holographic entanglement entropy,''
  JHEP {\bf 1311}, 074 (2013)
  
\bibitem{DHW} 
  X.~Dong, D.~Harlow and A.~C.~Wall,
  ``Reconstruction of Bulk Operators within the Entanglement Wedge in Gauge-Gravity Duality,''
  Phys.\ Rev.\ Lett.\  {\bf 117}, no. 2, 021601 (2016)
  
\bibitem{tadashitmd} 
  A.~Del Campo and T.~Takayanagi,
  ``Decoherence in Conformal Field Theory,''
  JHEP {\bf 2002}, 170 (2020),
 arXiv:1911.07861.
  

\bibitem{KP} 
  A.~Kitaev, J.~Preskill,
  ``Topological entanglement entropy,''
  Phys.\ Rev.\ Lett.\  {\bf 96}, 110404 (2006);  M.~Levin, X.-G.~Wen,
  ``Detecting Topological Order in a Ground State Wave Function,''
  Phys.\ Rev.\ Lett.\  {\bf 96}, 110405 (2006);  S.~Dong, E.~Fradkin, R.~G.~Leigh and S.~Nowling,
  ``Topological Entanglement Entropy in CS Theories and Quantum Hall Fluids,''
  JHEP {\bf 0805}, 016 (2008)

\bibitem{LH} 
  L.~McGough and H.~Verlinde,
  ``Bekenstein-Hawking Entropy as Topological Entanglement Entropy,''
  JHEP {\bf 1311}, 208 (2013),
  [arXiv:1308.2342 ].


  \bibitem{Happy} 
  F.~Pastawski, B.~Yoshida, D.~Harlow and J.~Preskill,
  ``Holographic quantum error-correcting codes: Toy models for the bulk/boundary correspondence,''
  JHEP {\bf 1506}, 149 (2015)
  
\bibitem{zaletel}
M.P. Zaletel and R.S.K. Mong, ``Exact matrix product states for quantum Hall wave functions,"
Phys. Rev. B 86, 245305, (2012)

  \bibitem{Coleman} 
  S.~R.~Coleman,
  ``Why There Is Nothing Rather Than Something: A Theory of the Cosmological Constant,''
  Nucl.\ Phys.\ B {\bf 310}, 643 (1988).
  
  
\bibitem{DonHenry} 
  D.~Marolf and H.~Maxfield,
  ``Transcending the ensemble: baby universes, spacetime wormholes, and the order and disorder of black hole information,''
  arXiv:2002.08950. 
  


  \bibitem{FW} 
  A.~Almheiri, D.~Marolf, J.~Polchinski and J.~Sully,
  ``Black Holes: Complementarity or Firewalls?,''
  JHEP {\bf 1302}, 062 (2013)

 
  \bibitem{EngelhardtWall} 
  N.~Engelhardt and A.~C.~Wall,
  ``Quantum Extremal Surfaces: Holographic Entanglement Entropy beyond the Classical Regime,''
  JHEP {\bf 1501}, 073 (2015).
\bibitem{complexity} 
  A.~R.~Brown, D.~A.~Roberts, L.~Susskind, B.~Swingle and Y.~Zhao,
  ``Holographic Complexity Equals Bulk Action?,''
  Phys.\ Rev.\ Lett.\  {\bf 116}, no. 19, 191301 (2016)
\bibitem{formfactor} 
  J.~S.~Cotler {\it et al.},
  ``Black Holes and Random Matrices,''
  JHEP {\bf 1705}, 118 (2017)
  Erratum: [JHEP {\bf 1809}, 002 (2018)

  
    \bibitem{QKL}
X.L. Qi, H. Katsura, A.W.W. Ludwig, "General Relationship between the Entanglement Spectrum and the Edge State Spectrum of Topological Quantum States,"
Phys. Rev. Lett. 108, 196402, (2012)

\bibitem{shinsei} 
  X.~Wen, S.~Matsuura and S.~Ryu,
  ``Edge theory approach to topological entanglement entropy, mutual information and entanglement negativity in Chern-Simons theories,''
  Phys.\ Rev.\ B {\bf 93}, no. 24, 245140 (2016),
  

\bibitem{cardy-calabrese} 
  P.~Calabrese and J.~L.~Cardy,
  ``Time-dependence of correlation functions following a quantum quench,''
  Phys.\ Rev.\ Lett.\  {\bf 96}, 136801 (2006)
  

  \end{thebibliography}
  \end{document}